\documentclass[12pt]{spieman}  
\usepackage{amsmath,amsfonts,amssymb}
\usepackage{graphicx}
\usepackage{setspace}
\usepackage{tocloft}
\usepackage[left, modulo]{lineno}

\title{Starshade formation flying I: optical sensing}

\author[a, b* ]{Michael Bottom}
\author[a]{Stefan Martin}
\author[a]{Eric Cady}
\author[a,c]{Megan C. Davis}
\author[a]{Thibault Flinois}
\author[a]{Dan Scharf}
\author[a]{Carl Seubert}
\author[a]{Shannon Kian Zareh}
\author[a]{Stuart Shaklan}
\affil[a]{Jet Propulsion Laboratory, California Institute of Technology, 4800 Oak Grove Dr, Pasadena, CA 91109}
\affil[b]{Institute for Astronomy, University of Hawaii, 640 N. Aohoku Pl, Hilo, HI 96720}
\affil[c]{Michigan State University, 220 Trowbridge Rd, East Lansing, Michigan 48824}

\newcommand{\specialcell}[2][c]{%
  \begin{tabular}[#1]{@{}c@{}}#2\end{tabular}}
\newcommand{\mytilde}{\raise.17ex\hbox{$\scriptstyle\mathtt{\sim}$}}

\cftpagenumbersoff{figure}
\cftpagenumbersoff{table} 
\begin{document} 
\maketitle

\begin{abstract}
A key challenge for starshades is formation flying.  To successfully image exoplanets, the telescope boresight and starshade must be aligned to approximately one meter at separations of tens of thousands of kilometers.  This challenge has two parts: first, the relative position of the starshade with respect to the telescope must be sensed; second, sensor measurements must be combined with a control law to keep the two spacecraft aligned in the presence of gravitational and other disturbances.  In this work, we present an optical sensing approach using a pupil imaging camera in a 2.4-meter telescope that can measure the relative spacecraft bearing to a few centimeters in one second, much faster than any relevant dynamical disturbances.  A companion paper will describe how this sensor can be combined with a control law to keep the two spacecraft aligned with minimal interruptions to science observations.
\end{abstract}


{\noindent \footnotesize\textbf{*}Corresponding author,  \linkable{mbottom@hawaii.edu} }

\begin{spacing}{2}   
\section{Introduction}
\label{sect:intro}  
Starshades, large occulters designed to artificially block starlight, offer a path to imaging and spectroscopy of Earth-like extrasolar planets.  The carefully shaped petals of a starshade create a dark stellar eclipse over the entrance pupil of a space telescope by controlling diffraction so that starlight does not concentrate along the optical axis as in the ``Arago spot'' phenomenon.  Current mission concepts envision 20-100 meter starshades positioned tens of thousands of kilometers in front of their respective space telescopes.\cite{ seager2019starshade, gaudi2019habitable}

A key challenge in the starshade concept is formation flying, as the starshade shadow is only sufficiently dark in a region 1-2 meters wider than the pupil of the telescope.\footnote{This is intentional, as creating a much wider shadow requires a much larger starshade, which blocks more of the inner orbits surrounding the star.  It is also harder to build and launch.}  With this sized shadow, the relative bearing, or ``shear,'' between the starshade and telescope must be maintained to approximately one meter, despite the large distances between the spacecraft.\cite{cash2011analytic}  (The separation tolerance is far less stringent, at about 250 km).  Formation flying has two components, sensing and control; sensing for determining the position of the starshade with respect to the telescope, and control for using the sensor data and onboard thrusters to efficiently maintain the required flight tolerances in the space environment.  This paper will address the challenge of sensing, and a companion paper\cite{flinois_inprep} will address the challenge of control.

A brief review of the mission concept is in order.  The starshade is intended to work with a space telescope, and must be launched and deployed.  The 20-30 meter optic is rolled up into a cylinder to fit in a rocket fairing, then launched into space.  In space, it separates and unfurls, with petal shape tolerances of \mytilde{}100 $\mu$m and petal position tolerances of \mytilde{}1 mm being required.\cite{webb2014successful}  Once unfurled, the starshade uses thrusters to maneuver itself between the target star and space telescope, with typical spacecraft separations of tens of thousands of kilometers.  When aligned, science observations occur, after which the starshade moves to the next target star.  These retargeting maneuvers can take days to weeks, as tens of degrees between target stars translate into slews of hundreds of thousands of kilometers.

The pointing and acquisition problem can be divided into three different regimes, which we refer to as coarse, medium, and fine.  In the coarse and medium regime, the relative starshade position is determined through measurements of distance and angle.  Distance can be determined by a time-of-flight S-band radio link between the spacecraft, with an accuracy of 500 meters, which is a negligible error given the 250 km range tolerance.  In the coarse regime, at relative separations of less than 600 kilometers from target, a wide-angle (\mytilde{}3$^{\rm o}$) laser beacon on the starshade can be used in conjunction with an external star camera.  Here, the angles between the starshade (which may be identified as a blinking or uncatalogued point source) and a set of reference stars may be measured with a star tracker camera to better than 2 arcseconds,\cite{liebe1995star} corresponding to a shear accuracy of 400 meters for a typical separation of 40,000 km.  A switch to the medium-sensing mode occurs once the angular separation from the target star is less than a few arcseconds.  Here, the telescope's internal science camera is used to sense the starshade position, using the same concept of differential measurement of point-spread functions, but with a much finer accuracy of 20 milliarcseconds, or about 4 meters.  This is maintained until the starshade begins to occult the star, at a separation of about 15 meters.  At this point, the two point-spread functions are no longer well separated in the science camera, and a switch occurs to the fine-sensing mode.\footnote{In flight, we anticipate the possibility of a brief transition period without good sensor measurements between the medium and fine sensing mode, from 30 meters out to 10 meters out.  This will require some knowledge of the starshade position and velocity to plan the rocket firings.\cite{scharf2016precision}}

\begin{figure}[!htb]
  \centering
  \begin{minipage}[b]{0.95\textwidth}
    \includegraphics[width=\textwidth]{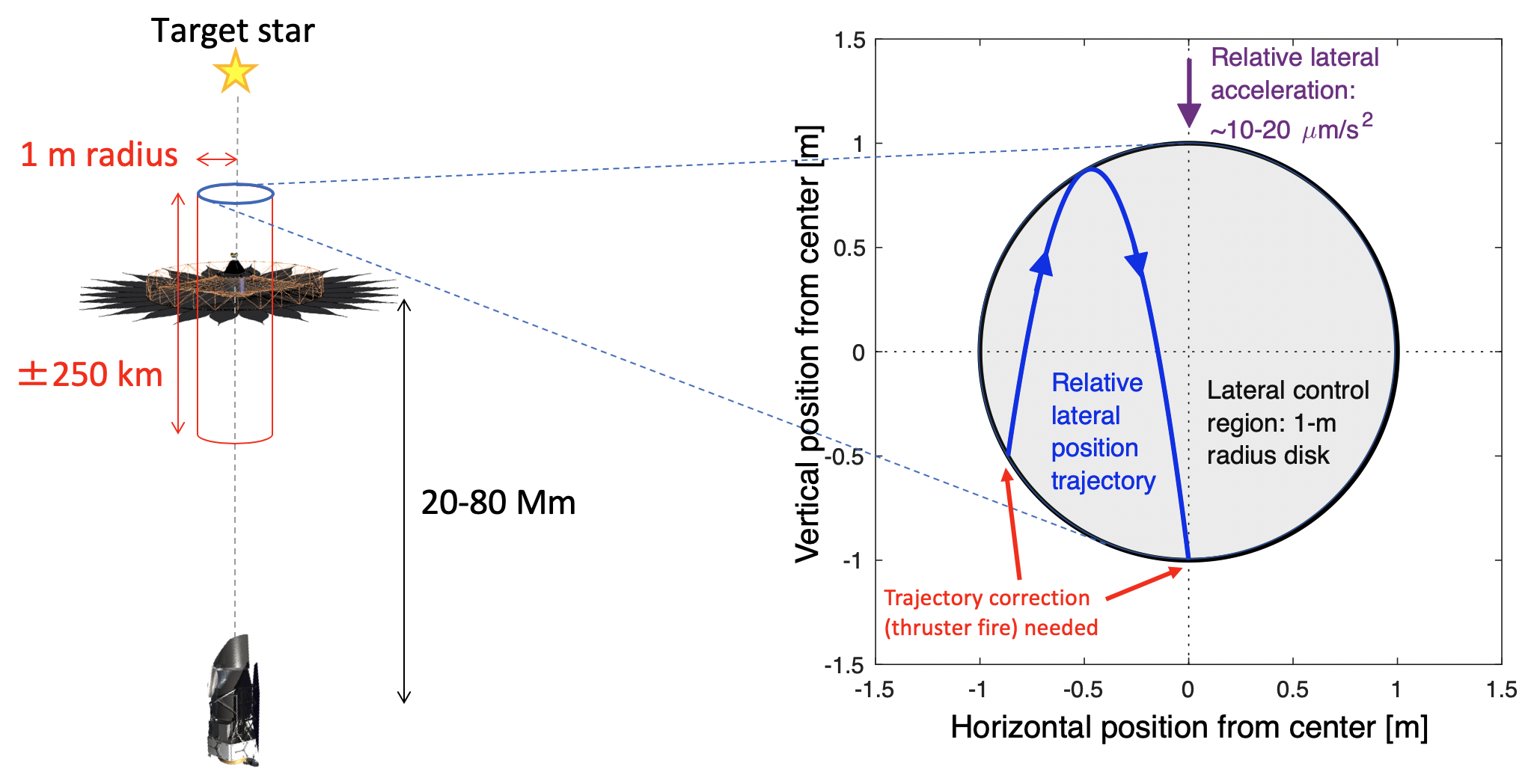}
    \caption{ \label{trajectory}Outline of the formation flying problem.  The 20-80 Mm position between the telescope and starshade must be maintained in a cylinder of 1 m in radius and 250 km in length.  The radial control consists of firing the starshade's thrusters to execute ballistic trajectories within the 1 m deadband, to counteract gravitational accelerations of 10-20 $\mu$m/s$^2$.}
       \end{minipage}
\end{figure}

The fine-sensing mode, for stationkeeping during science operations, uses an internal pupil imager of the telescope to determine the relative shear between it and the starshade.  During stationkeeping, the relative bearing between the spacecraft and starshade must stay within the 1 meter deadband, with trajectory corrections required every \mytilde{}10 minutes to compensate for the $\mu$g differential gravitational force and solar radiation pressure.  These corrections are accomplished through starshade thruster firings, which cause ballistic trajectories within the deadband.  The overall speed of these maneuvers is fairly slow, with maximum velocities of \mytilde{}2 cm/s at the deadband boundary.

\section{Optical sensing}
The 1 meter stationkeeping demand on a starshade mission leads to sensing needs that are more stringent than 1 meter.  A starshade technology maturation program called \textit{S5}\cite{willems2018starshade} defined the sensing requirements such that the 3-$\sigma$ error signal would be less than 30 cm, for a \mytilde{}2.4 m telescope aperture.  While no requirement on sensing cadence was given, the rate must be high enough as to permit control within the 1 meter deadband under the disturbances experienced in flight.                    

\begin{figure}[!htb]
  \centering
  \begin{minipage}[b]{0.65\textwidth}
    \includegraphics[width=\textwidth]{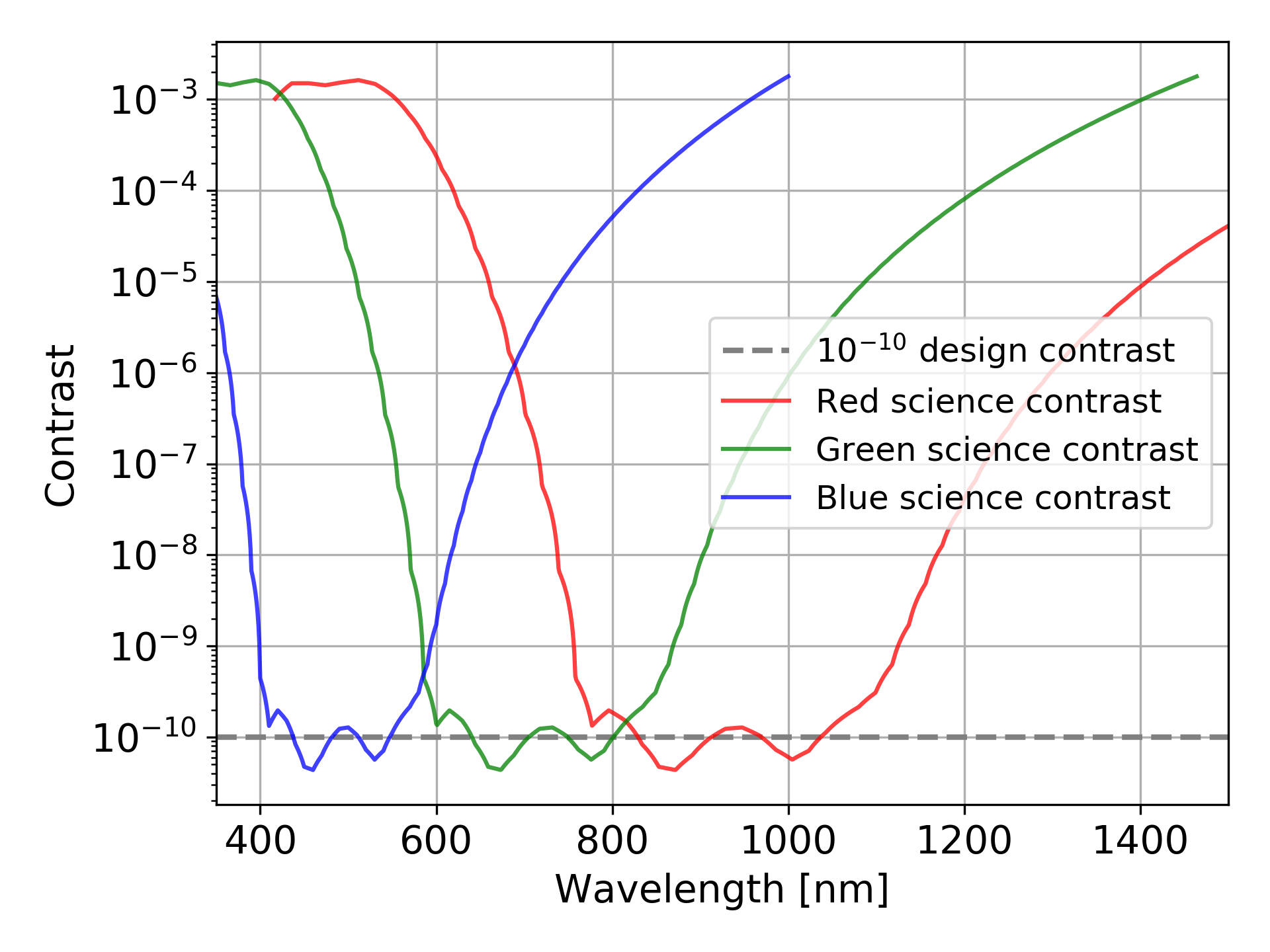}
    \caption{ \label{science_bands}Plot of the starshade suppression for the red, green, and blue science bands.}
       \end{minipage}
\end{figure}

Starshades are designed to operate in the Fresnel regime, where the dimensionless ``Fresnel number'' $F \sim r^2/(\lambda Z)$ is in an intermediate regime of $F>10$.  Here $r$ is the starshade radius, $\lambda$ the wavelength, and $Z$ the separation between the starshade and telescope, and optical propagation physics is preserved when the Fresnel number is the same.  Spectroscopy of exoplanets requires rather broad wavelength coverage to measure different spectral features.  Interesting molecular signatures exist from 400-800 nm, including from oxygen, ozone, and water, and in the case of the Earth, the ``red edge'' at 700nm pointing to the presence of vegetation.  However, moving from 400-800 nm changes the Fresnel number by of a factor of two, which is challenging to accommodate in the optical design.  This leads to the definition of  ``science bands'' corresponding to different spectral regions, where the starshade must change its distance from the telescope, so that $\lambda Z \sim \mathrm{constant}$.  For example, in the case of the WFIRST starshade concept, the blue science band is from 400-600 nm, at a separation of about 40,000 km; the green from 600-800 nm, at separations of 30,000 km; and the red at 800-1000 nm, at separations of 20,000 km (Figure \ref{science_bands}).

Starshade suppression rapidly degrades when operated outside of the designed science band (ibid).  For example, for the green science band, the starlight suppression at 600 nm is nearly 10$^{7}$ times higher than at 500 nm, despite these wavelengths being only 100 nm apart.  This bright ``leaked'' light must be blocked internally in the telescope using bandstop filters.  Note that the starshade only suppresses the on-axis starlight.  Off-axis planet light is unaffected.  

\begin{figure}[!htb]

  \centering
  \begin{minipage}[b]{1.0\textwidth}
    \includegraphics[width=\textwidth]{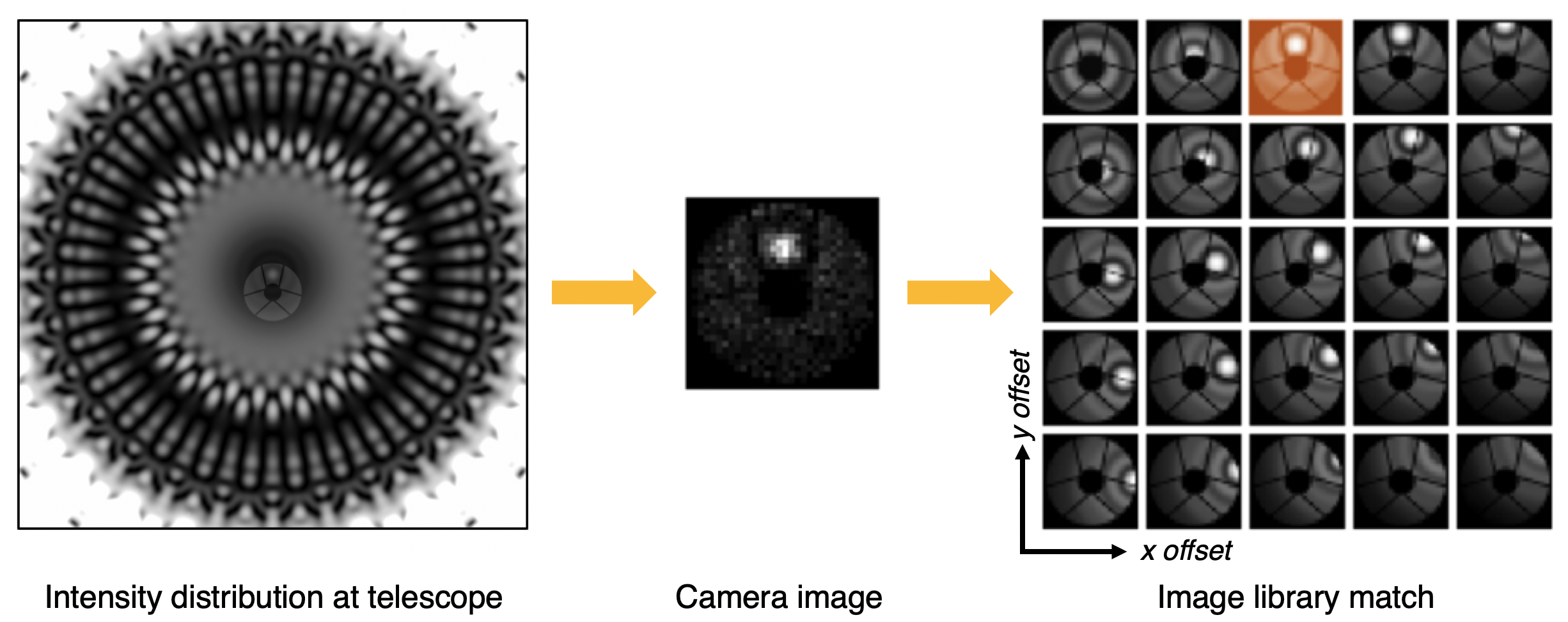}
    \caption{ \label{image_library}(Left) 20x20 meter image of out-of-band light pattern at the telescope pupil, stretched to show detail.  A transparent image of the telescope pupil at the same scale is overlaid. (Center) A simulated image of the telescope's internal pupil camera, showing the Arago spot slightly offset from center.  (Right) the noisy pupil image is compared against a library of precomputed images with known shear offsets, and the best match corresponds to the relative shear of the starshade and telescope, in this case at shear offset (0.0, 0.6)}
       \end{minipage}

\end{figure}

It is this out-of-band stellar ``leakage'' that is actually key to sensing the shear.  The starshade cannot effectively suppress light at these wavelengths, and it focuses behind the starshade as a bright core of light, similarly to the classical spot of Arago.  At these intermediate distances and Fresnel numbers, the spot width is on the order of tens of centimeters, surrounded by a dark ring and and complex diffraction artifacts due to the petals (Figure \ref{image_library}).  While the light distribution does not meaningfully change with starshade distance deltas of hundreds of kilometers (the Fresnel number stays almost the same), it precisely tracks the shear offset of the starshade: if the starshade moves vertically by 25 cm, the pattern will move vertically by that same amount.  Thus, the offset of the spot and surrounding light with respect to the center of the telescope pupil can be used to determine the shear between the starshade and the telescope.

In order to effectively sense this signal, some way of measuring the light distribution in the pupil of the telescope is needed.  (Analyzing this light in the focal plane is much less effective, as shear changes lead to only subtle differences in the point-spread function.)  Pupil sensors, which directly image this light distribution, are not common in space telescopes, as they either require a movable optic to shift the focus to the position of the pupil, or a separate camera.  However, they are critical elements in high contrast imaging systems, where they can be combined with interferometric elements or lenslets to directly sense internal wavefront aberrations, which can then be corrected by deformable mirrors.  Typically, the pupil is imaged onto a small-format detector, for high readout speeds.  A familiar example of a pupil sensor is the Shack-Hartmann wavefront sensor.

Previous work has examined the viability of optical sensing to determine the shear of the starshade with respect to the telescope.  Noecker 2007 \cite{noecker2007alignment} reviewed some potential methods of sensing the relative shear, and introduced the concept of using pupil sensing using a set of outrigger telescopes on ``booms'' to measure the relative gradient of light.  A similar implementation was proposed in Sirbu 2011,\cite{sirbualignment} which had an internal infrared octant sensor to provide the sensing information.  Harness and Cash 2015 developed an analytic method of centroiding using pupil plane images that is similar to the method proposed in this work.\cite{harness2015enabling}  Image plane sensing is also possible; Scharf et al. 2016 \cite{scharf2016precision} presented a method using the science camera, by using difference images of the starshade (and its laser beacon) and target star.  Image plane sensing is more challenging to implement, given the small angles involved, and has an expected measurement precisions of 1 meter (3-$\sigma$). (ibid)  It is not able to reach the performance of pupil-plane schemes.

The work presented in this paper builds on the previous developments, with the goal of characterizing the precision of pupil plane sensing with realistic inputs for the radiometric error budget, assuming a WFIRST-sized  telescope and the pupil camera approach of Harness and Cash.  We present analytic calculations and detailed numerical simulations of the expected sensing performance and validate them against laboratory experiments performed at similar Fresnel numbers and signal-to-noise ratios.  Unlike the analytic spot centroiding algorithm of Harness and Cash, we develop a sensing algorithm that is based on image matching, where the data from the pupil camera is compared against a library of pre-computed pupil images corresponding to different starshade offsets; see Figure \ref{image_library}.  This is a brute-force algorithm, but is tractable given the minimal degrees of freedom in the problem (just two, the offset in the horizontal and vertical direction).  This is essentially equivalent to matched filtering, and as such, should be optimal, with uncertainties in position driven by photon and detector noise, and not imperfections in the matching algorithm.  While the algorithm appears to be easily manageable on a flight computer, simpler algorithms (such as gradient-based centroiding) could be developed that have a much lower memory footprint and faster speed.

Portions of the text below (particularly describing experimental setup and design) are repeated verbatim from a previous conference proceeding \cite{bottom2017precise} and the technical report from NASA's S5 program describing the formation flying milestone.\cite{nasas5}

\subsection{Radiometry}
The amount of photons detected by the pupil sensor depends on the star brightness, the starshade suppression, the internal optical efficiency of the telescope optics up to the sensor, and the detector quantum efficiency.  All of these terms have a dependence on wavelength, of course.  

\subsubsection{Stellar input flux}
For the stellar models, we used a solar type model from the ATLAS9 library\cite{castelli2003modelling} (T$_{\rm eff}$=5750K, log[g] = 4.5, [Fe/H]=0).  We validated the stellar spectrum code against a standard measured solar irradiance spectrum (ASTM E-490), finding agreement in spectral flux density at the few percent level from 300-1000 nm.  Validation against filter photometric zeropoints also agreed to within a few percent.   The stellar spectral type is a minor contributor to the photon budget, as this causes variations in flux by at most a factor of \mytilde{}3 for stars of the same $M_{V}$ at any of the relevant wavelengths.  The main contributor to the stellar photon budget is the apparent magnitudes of the target stars, which range from $M_{V}$=-1.5 to 5.3, a factor of about 500.

\subsubsection{Starshade optical transmission}
The starshade optical transmission plays the largest role in the overall photon budget, as it varies by six to seven orders of magnitude between the science bands (deep suppression at 10$^{-10}$) and in the sensing bands (10$^{-3}$ to 10$^{-4}$), with a steep transition with wavelength of $C\propto \lambda^x$, where $x$\mytilde{}15, as shown in Figure \ref{science_bands}.  Starshade optical models have been validated at better than the 10$^{-10}$ level in laboratory demonstrations (Harness et al., in prep).  In this work, we are primarily interested in the performance at 10$^{-3}$ to 10$^{-4}$, where optical modeling uncertainties are negligible.

\subsubsection{Telescope efficiency}

\begin{figure}[!htb]
  \centering
  \begin{minipage}[b]{0.95\textwidth}
    \includegraphics[width=\textwidth]{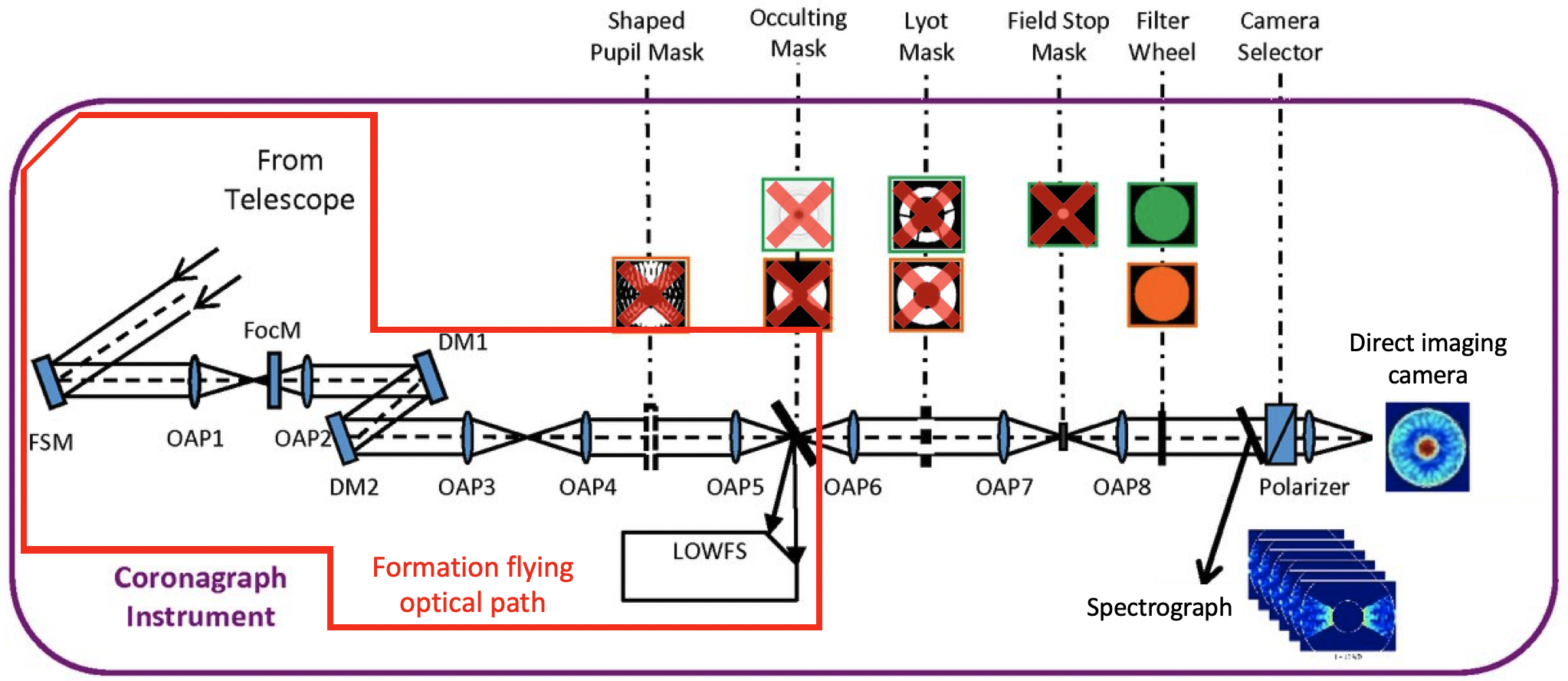}
    \caption{ \label{wfirst_ff} The formation flying path of the WFIRST coronagraph, which uses all the optics up to the low-order wavefront sensor.  Red x's refer to optics moved out of beam for starshade operations.  Figure adapted from Tang et al. 2017 \cite{tang2017wfirst}}
       \end{minipage}
\end{figure}

In the case of starshade operations with the WFIRST coronagraph, the optical train does not use any of the complex masks or stops that make the coronagraph so effective at suppressing starlight.  These are simply moved out of the way, leading to a much higher throughput on the science and spectrograph cameras.  The deformable mirrors and other adaptive optics are also not actively controlled, but set to predetermined ``flat'' setpoints.\footnote{A ``rough'' correction of 10 nms wavefront error, which would be unacceptable for coronagraphic performance, is still better than 98\% Strehl ratio, so has a negligible effect on starshade planet sensitivity}   There are about 20 optical surfaces between the telescope pupil and the pupil sensor (Figure \ref{wfirst_ff}), which is not uncommon for a coronagraph, but leads to throughput losses compared to a purpose-built pupil imaging camera.  Additionally, while it has not yet been decided what kind of filter (if any) will be present in front of the pupil camera (the low-order wavefront sensor, or LOWFS) for starshade operations, we chose to limit the spectra to broad bandpasses that corresponded to the peak out-of-band light for the different science modes (see Figure \ref{science_bands}).  These correspond to 400-540 nm for the red band, 400-435 for the green, and 870-900 for the blue.  Operating much outside than these bandwidths will not  have a large effect, as the starshade suppression increases dramatically with wavelength.

\begin{figure}[!htb]
  \centering
  \begin{minipage}[b]{0.65\textwidth}
    \includegraphics[width=\textwidth]{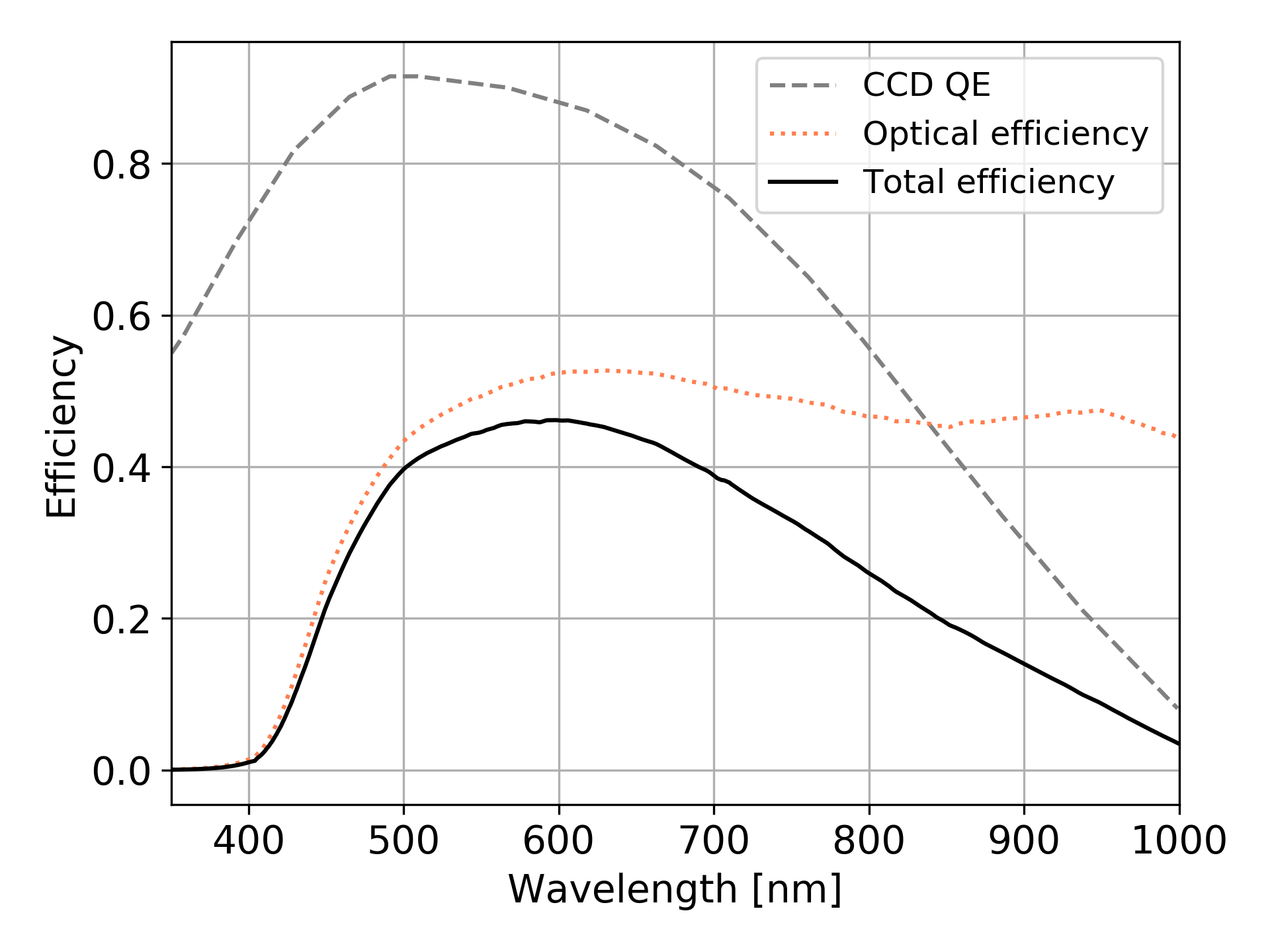}
    \caption{ \label{efficiency} Plots of the CCD quantum efficiency, optical efficiency, and combined efficiency of the WFIRST coronagraph (the input to our radiometric model)}
       \end{minipage}
\end{figure}

To compute the total efficiency to the LOWFS camera, we combined transmission curves of the individual optical elements from the WFIRST coronagraph optical design (Hong Tang, priv. comm.), and the camera quantum efficiency from the manufacturer data sheet (CCD201 from Teledyne e2V).\cite{morrissey2018photon}  Figure \ref{efficiency} shows the resulting curves, with the total efficiency not exceeding 50\% anywhere in the relevant bandpass.   The wavelength cutoffs are set by aluminum and silver reflectance at the blue end and detector quantum efficiency at the red end.  While the optical design is not finalized, it is not the driving factor for sensing performance, as we will show that changes in throughput of a factor of two will not preclude accurate sensing.

\subsection{Analytic calculations}
It is possible to get a rough estimate of the sensing performance using simple scaling arguments, in particular the ``centroid accuracy'' formula,
\begin{equation}
 \sigma_x = \frac{\rm{FWHM}}{c \cdot \rm{SNR}}
 \label{centroid_formula}
 \end{equation}
 where $\sigma_x$ is the spot centroid accuracy (1$\sigma$), FWHM is the spot full-width at half maximum, SNR is the spot signal-to-noise ratio, and $c$ is a constant of order unity that depends on the exact morphology of the PSF.  This formula is used in astrometry,\cite{king1983accuracy, kaiser2000new} with a value of $c=2$ being appropriate for Gaussian or Moffat-like stellar profiles.

We calculate the width of the Arago spot FWHM for the numerator using an analytic approach assuming the starshade is circular (the petals suppress contrast, but have a minor effect on spot size).  In this case, the intensity distribution near the optical axis has a functional form of 

\begin{align}
I(x, y)&\propto J_0^2\left( \frac{2 \pi (x^2 + y^2)^{1/2} r}{\lambda z}\right)
\end{align}

where  $x, y$ are the coordinates at the pupil of the telescope; $J_0$ refers to the zeroth Bessel function of the first kind; and $r, \lambda, z$ are the radius of the starshade, wavelength of light, and distance from the starshade to the telescope. The FWHM of this distribution is:

\begin{equation}
\mathrm{FWHM} \approx \lambda z/(\pi r)
\label{eqn_fwhm}
\end{equation}

What remains to evaluate in Equation \ref{centroid_formula} is the signal-to-noise ratio. For the SNR, we use the ``CCD formula,'' 
\begin{equation}
SNR = \frac{N_{ph}}{\sqrt{N_{ph} + n_{ap}\rm{RN}^{2}}}
\end{equation}
where $N_{ph}$ is the number of photons in the spot, $n_{ap}$ is the number of pixels in covered by the spot, and $RN$ is the readout noise per pixel, conservatively assumed to be 5 electrons.  This assumes readout noise and photon noise dominate the error budget, neglecting terms relating to dark current and sky background, which are not large with the baselined EMCCD detector (the e2v CCD201) in a space environment.  For the number of photons in the spot $N_{ph}$, we assume all the photons at the pupil (multiplied by the system efficiency) end up in a spot with a FWHM given by Equation \ref{eqn_fwhm}, rather than computing the numerical diffraction pattern and considering the pupil obscurations and numerical propagation through the telescope optics.

\begin{figure}[!htb]
  \centering
  \begin{minipage}[b]{0.65\textwidth}
    \includegraphics[width=\textwidth]{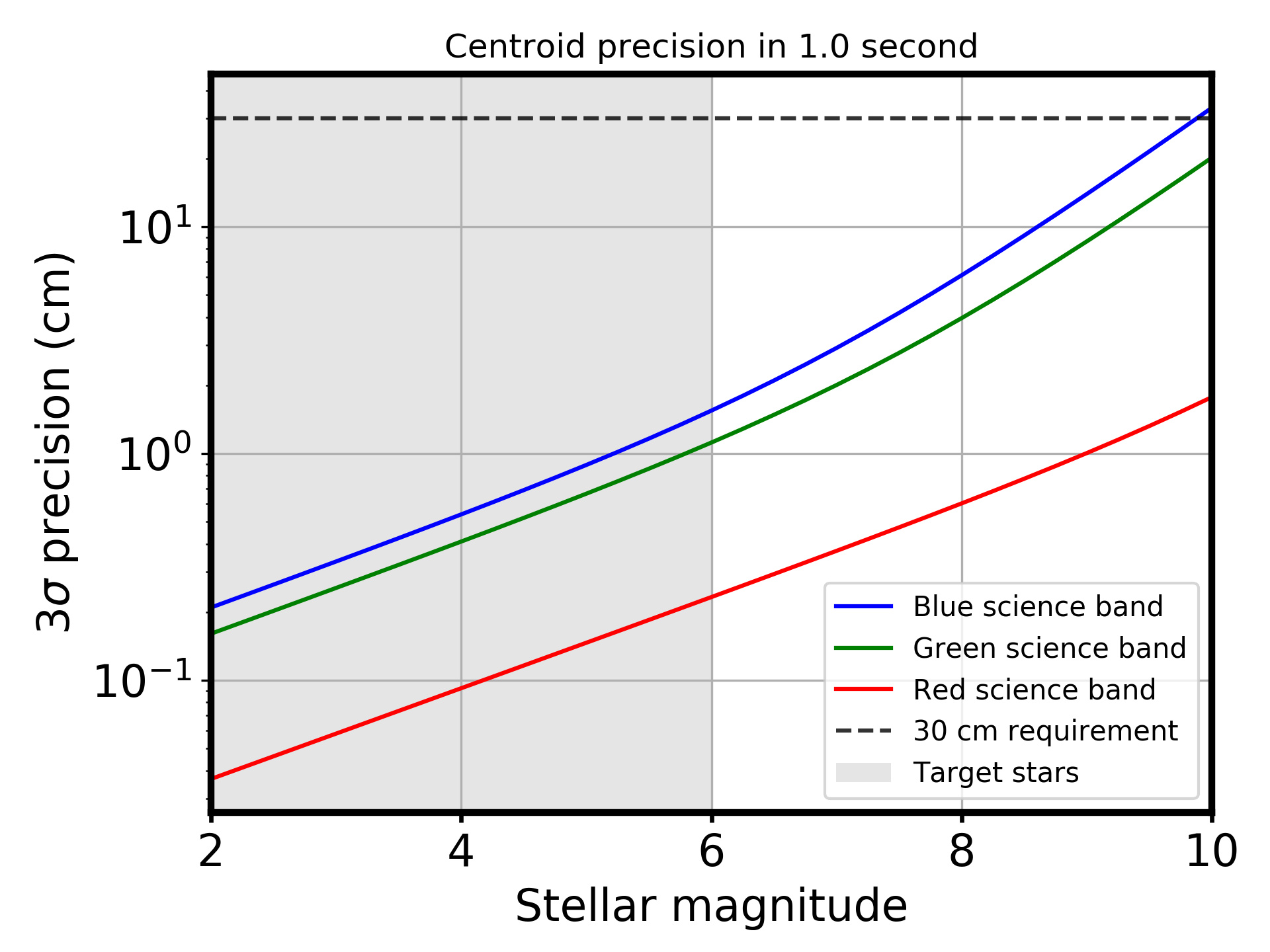}
    \caption{ \label{analytic_precision}Analytic calculations of centroid accuracy for the red, green, and blue science band.  For typical starshade target stars of $M_V<6$, the accuracy easily exceeds the 30 cm, 3-$\sigma$ requirement in 1 second of exposure time.}
       \end{minipage}
\end{figure}

Figure \ref{analytic_precision} shows the results of the analytic predictions.  For typical target star brightnesses of $M_V<6$, the predicted accuracy in one second of exposure time is better than 3 cm in all science bands, easily exceeding the 30 cm requirement.  The 30 cm requirement only becomes significant at around 10th magnitude, in the blue and green science bands.  This implies that with a suitably designed sensing algorithm, sensing accuracy should not be a challenge for starshade operations.

This is a simple approximation of the achievable precision, considering only the starshade size, the telescope efficiency, and the detector sampling and noise.  However, it gives a useful estimate of the performance which may be expected, and the scaling.  As we will later show, it agrees well with the numerical results, supporting the validity of the more detailed simulations.

\section{Numerical Simulations}
The analytic estimates of performance did not refer to a particular centroiding algorithm, but considered theoretical limits based on spot size and signal-to-noise ratio.  The true images will not be simple spots, but will have extra structure, particularly when considering pupil obscurations.  A spot centroid algorithm works best with an unobscured spot, and would potentially fail if the spot exited the pupil area or was blocked by the secondary mirror.  These shortcomings can be fixed in principle, but this leads to a second issue, which is that any centroid algorithm will add some error to the spot position beyond the fundamental limits imposed by spot shape and photon noise.  We adopted a different approach that would avoid these obscuration and visibility issues, with minimal error added by the algorithm, as will be described below.

We constructed detailed numerical simulations to create accurate models of the images seen on the pupil camera, and to estimate the sensing performance.  These simulations used electric field propagation from the starshade through the telescope optics.  First, the electric field of the starshade was calculated using the boundary integral method of Cady 2012.\cite{cady2012boundary}  The electric field was then 1) shifted to account for the input offset positions, 2) multiplied by the telescope pupil aperture function to add the central obscuration, outer diameter and spiders, 3) propagated to the internal Zernike sensor plane, 4) multiplied by the Zernike phase function, 5) propagated to the low-order wavefront sensor camera, and 6) converted to an intensity.  We used Fourier transforms to propagate between focal and pupil planes, and performed all propagations using the intensity-weighted wavelength, reasonable given the limited sensing bandwidth.  This procedure was repeated on a 2 cm grid to build up the image library.

The Zernike sensor, an interferometric pupil phase sensor in the coronagraph, is only useful for low-order wavefront sensing.  For shear sensing, it serves no useful function.  Its effect on the shear signal is to create mild intensity gradients over the image, and actually diffract a good deal (10-20\%) of light outside the pupil imager, similar to a coronagraph.  The reason we included it is that it reduces the flux, making it a more conservative assumption, and it is expected to exist in the system in a baseline configuration.  In practice, it would be possible to use the Zernike sensor to sense tip/tilt separately from shear, but this is beyond the scope of this work.  Adding telescope tip/tilt jitter to the simulations had less than a 10\% effect on the sensing precision, but increased computation times signficantly, so we neglected to include it.  We also ignored motion blur from the starshade, for two reasons.  First, the maximum speed of 2 cm/s will contribute at most 2 cm of sensing error for a 1 second exposure time, which is well below the required sensitivity.  Furthermore, actual target stars are so bright that the exposure times will be much less than one second, so minimal motion blur will occur.

Here we introduce the particular algorithm to determine the shear offset (for a full description of the algorithm, including storage requirements and computational complexity, see the Appendix).  Rather than using a spot  centroid algorithm, we use a least-squares image matching procedure equivalent to a matched filter,  which should minimally contribute to the derived uncertainty in position.  Each input image $I$ is converted into a vector of $n$ pixels (for example, $n$ is 1024 for a 32x32 image), is normalized by the sum of the image intensities $I_m = I/ \sum_n I$ and the scalar $e^2_{x,y} = (L_{x,y}  - I_m)^2$ is calculated for each image $L_{x,y}$ in the library (all library images are also mean-subtracted).  The library image with the lowest $e^2_{x,y}$ is selected, and its position $x, y$ becomes the starshade position estimate.

\begin{figure}[!tbp]
  \centering
  \begin{minipage}[b]{0.45\textwidth}
    \includegraphics[width=\textwidth]{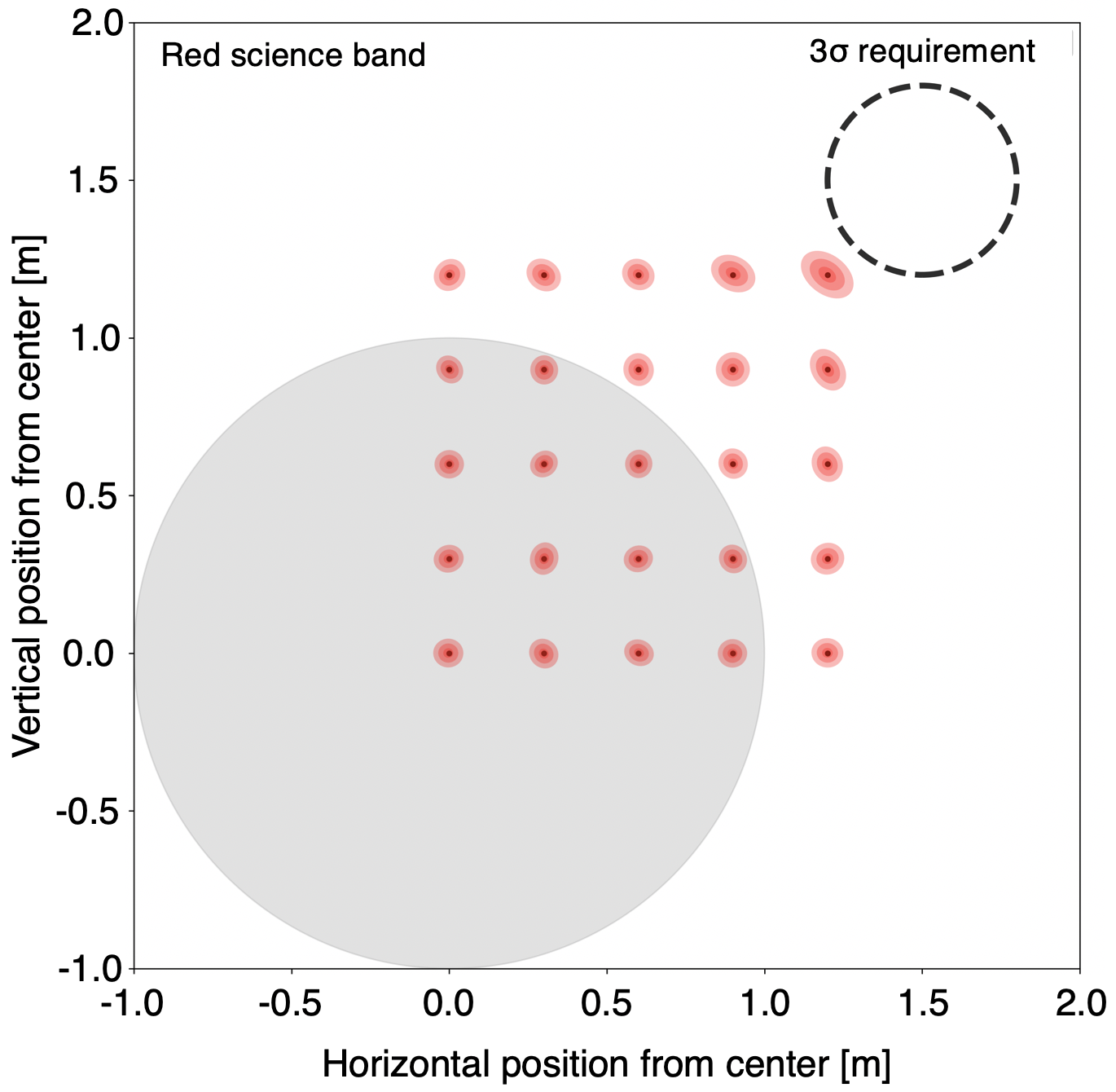}
  \end{minipage}
\begin{minipage}[b]{0.45\textwidth}
    \includegraphics[width=\textwidth]{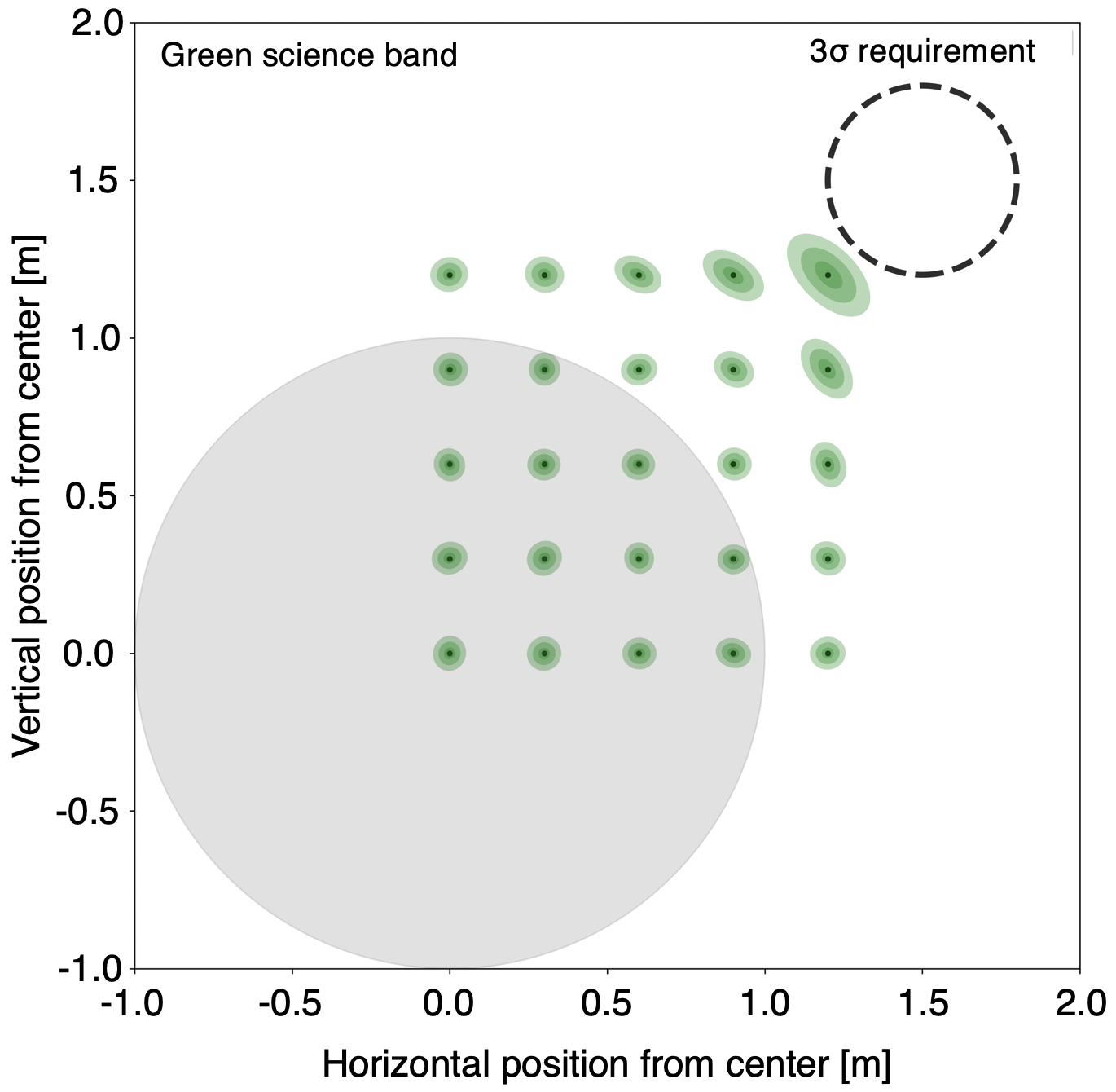}
  \end{minipage}
\begin{minipage}[b]{0.45\textwidth}
    \includegraphics[width=\textwidth]{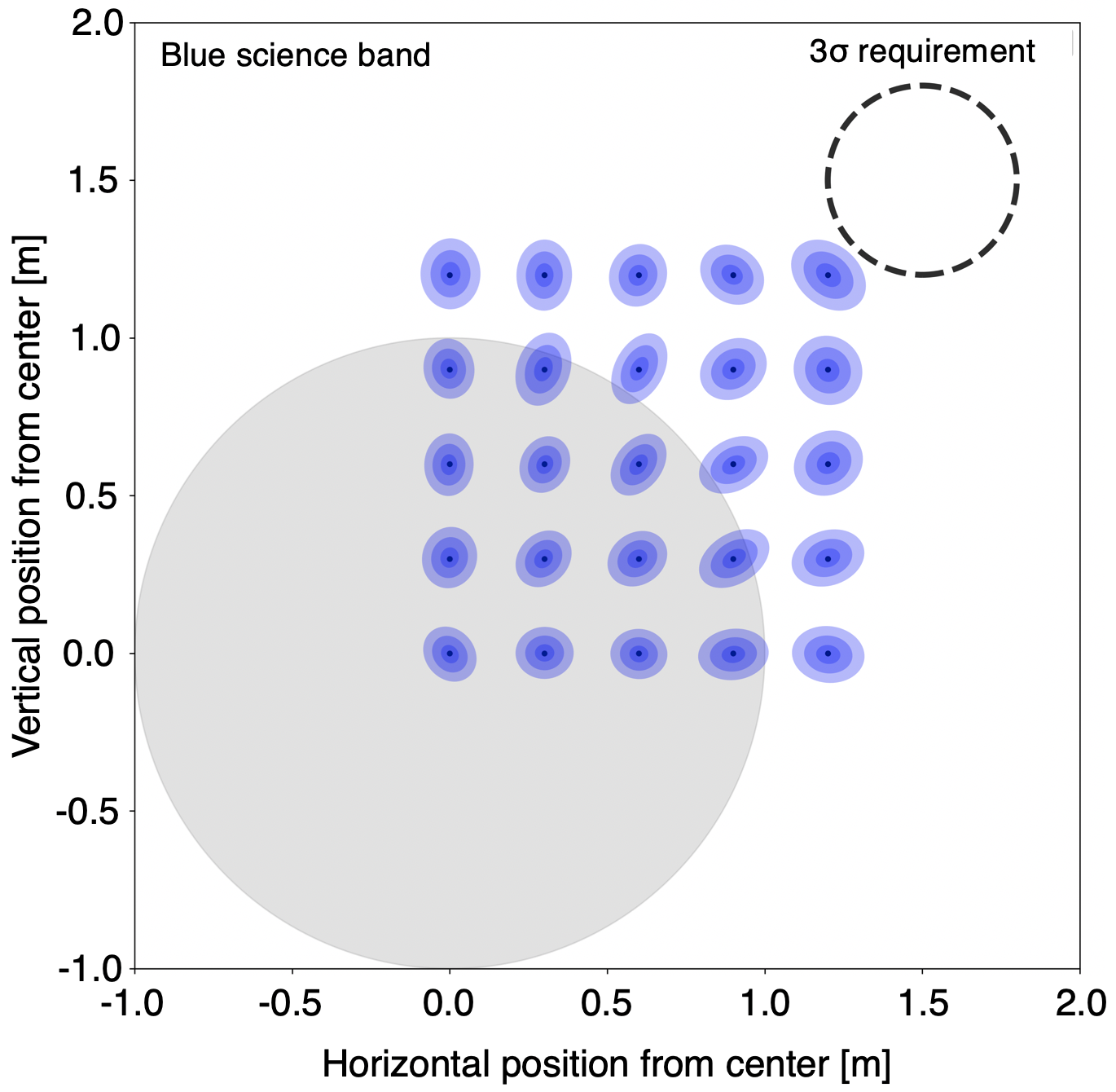}
  \end{minipage}
  
  \caption{\label{num_sim_images} Numerical simulations of red, green, and blue science bands for 10th, 8th, and 8th magnitude stars, respectively.  The grey circle shows the 1-meter control radius, and the colored ellipses show the 1, 2, and 3-sigma errors for different sensing positions in 1 second of integration time.  The dashed circle shows the 30 cm control requirement from the S5.}
\end{figure}

\begin{figure}[!tbp]
  \centering
  \begin{minipage}[b]{0.45\textwidth}
    \includegraphics[width=\textwidth]{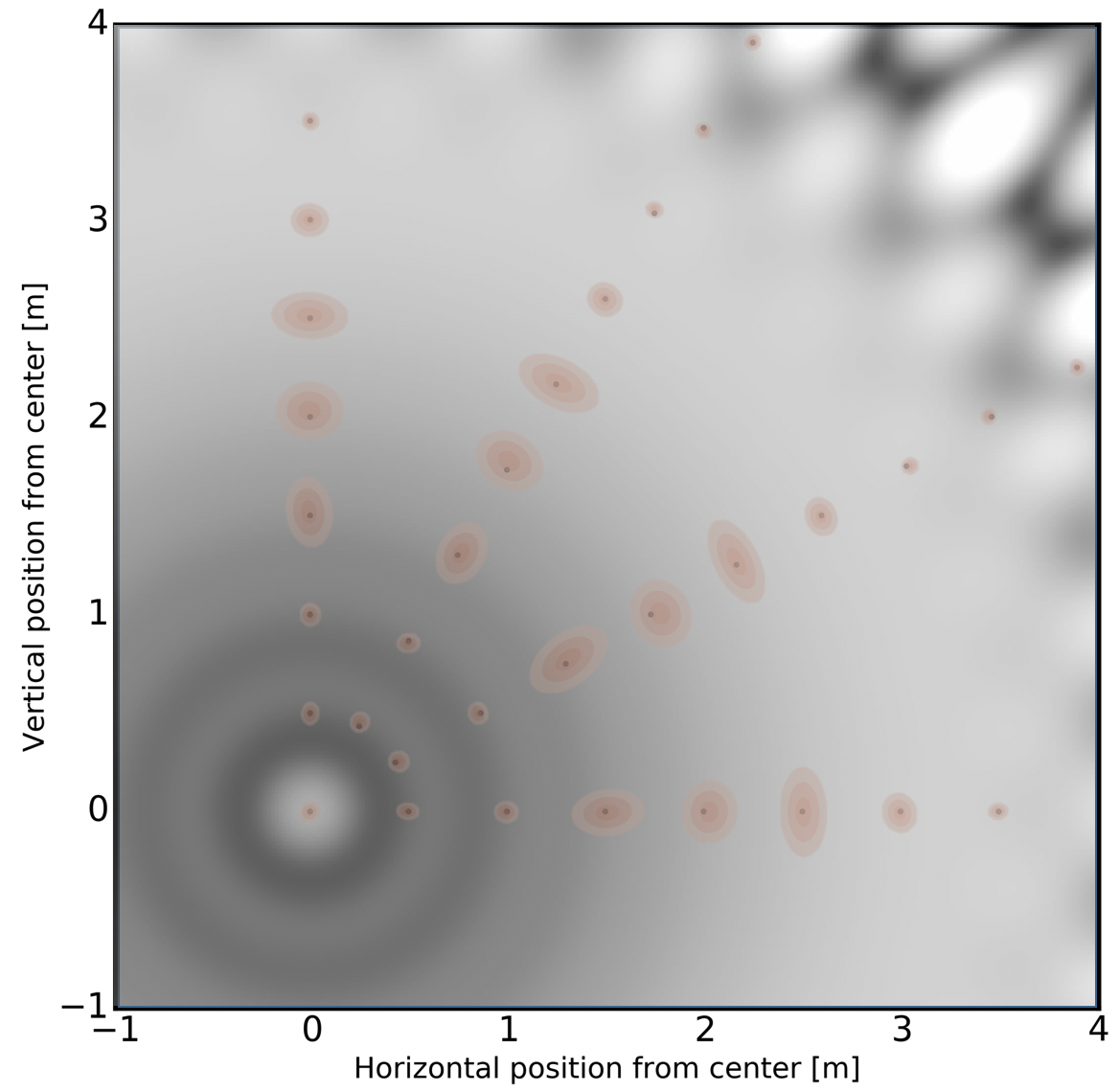}
  \end{minipage}
  \caption{\label{overplot_precision} Numerical computation of sensing precision, overlaid with the underlying light intensity.  The ellipses show the 1, 2, and 3 sigma contours in the red science band, for an eighth magnitude star, in a 4 meter region surrounding the central lobe.  The precision scales approximately as the second spatial derivative of the light intensity. }
\end{figure}

For the numerical simulations, we used library grid spacings of 2 cm calculated over the positive quadrant of a circle of less than 1.3 meters in radius.  We tested shear positions separated by 30 cm in one quadrant of the control region (the shadow is nearly perfectly circularly symmetric in the control region, and more complex diffraction effects do not appear until 4 meters from the center; see Figure \ref{overplot_precision}).  At each grid point, we used the same method to propagate the electric field through the starshade and telescope optics, but reduced the amplitude of the electric field to account for stellar magnitude and optical efficiency.  We generated 300 such realizations per point, added Poisson and readout noise, then matched them to the image library, saving the best-fit positions.  From these matches, we generated empirical 1, 2, and 3-sigma error ellipses,  shown in Figure \ref{num_sim_images}.

\begin{table}[htp]
\begin{center}
\resizebox{\columnwidth}{!}{%
\begin{tabular}{| l | c | c | c | c | c | c|}
\hline
\specialcell{Science\\ band} & \specialcell{Star\\ magnitude} &
 \specialcell{Flux density at pupil\\ (photons/m$^2$/s)}& 
 \specialcell{Science \\ wavelengths (nm)}& 
\specialcell{Guiding \\ wavelengths (nm) \\ $[$weighted$]$} & 
\specialcell{Median\\ numerical\\3-$\sigma$ error (cm)} & \specialcell{Analytic \\3-$\sigma$ error}\\
\hline
Blue 		& 8.0 	& 8200 	& 400-600		&870-1000 [937] 	& 9.7 & 6.1 \\
Green 	& 8.0 	& 2100 	& 600-800 	& 400-435 [424] 	& 3.6 & 3.9 \\
Red 		& 10.0 	& 11600 	& 800-1000 	& 400-540 [496] 	& 1.6 & 1.6 \\
\hline
\end{tabular}
}
\end{center}
\caption{\label{simulationtable} Table of numerical simulation results}
\end{table}%

The results of the numerical simulations are consistent with the earlier analytic estimates, with errors of a few centimeters being predicted at all science bands, with exposure times of 1 second, for stars of 8th to 10th magnitude (Table \ref{simulationtable}).  The largest discrepancy was the blue science band, with detailed numerical simulations being about 50\% worse than the analytic prediction.  In the blue science band, the guiding spot (in the near infrared) has the largest size (Eq. \ref{eqn_fwhm}), so that it is always partially obscured by the telescope's secondary mirror and supports.  This can be incorporated into the analytic estimate by  modified, spatially dependent values for the shape parameter $c$.  However, we used a fixed $c=2$ for an unobscured spot, so it is not surprising that the formula tends to predict better performance than the numerical simulations.  

To understand the shapes of the ellipses, we might expect that at each position, the precision will depend on the second derivative of the light intensity.  Constant intensity distributions will give no information as they are translation invariant.  Similarly, linearly increasing distributions will give no information due to the normalization by total flux.  However, higher orders will. \footnote{As a simple example, if you find yourself standing on a hill with shape $y= -|x|$, you can't tell where on the hill you are by examining the slope of the hill at your position.  But you can for the hill $y = -x^2$.}.  Said another way, the covariance matrix of the error ellipse will be inversely proportional to the Hessian of the light intensity.  This becomes more evident when plotting the sensing precision over a larger region, where there are both smooth and structured regions, as in Figure \ref{overplot_precision}.

\section{Laboratory Results}
\subsection{Overview}
A remaining question is whether the numerical simulations and analytic predictions give reasonable expectations for formation flying performance when compared against actual hardware.  We built the Starshade Lateral Alignment Testbed (SLATE) to validate the lateral position sensing approach in the lab.  The experimental design is the same as the numerical simulations, where the starshade is moved to a pre-determined offset, and the measured intensity on the camera is matched to a precomputed library of images using a least-squares algorithm.  This is repeated hundreds of times to determine the accuracy of the position matching.  In the following subsections, we give a description of the testbed design, hardware and optical considerations, and experimental results.

\subsubsection{Testbed design}
As previously mentioned, starshades are designed to operate at Fresnel numbers $F \sim r^2/(\lambda Z)$ of $<$20.  Here $r$ is the starshade radius, $\lambda$ the wavelength, and $Z$ the separation, and most optical propagation effects are preserved when the Fresnel number is the same.  Due to the large separation distances of $>$10000 km, it is not possible to optically validate a full-scale starshade on Earth.  However, aspects of a flight-like setup may be tested by quadratically decreasing distance $Z$ with starshade radius $r$.  Scaling down the starshade by a factor of one thousand requires a separation one million times smaller, allowing for optical validation with more manageable testbeds sized 1-100 meters.

\begin{figure}[!tbp]
  \centering
  \begin{minipage}[b]{0.65\textwidth}
    \includegraphics[width=\textwidth]{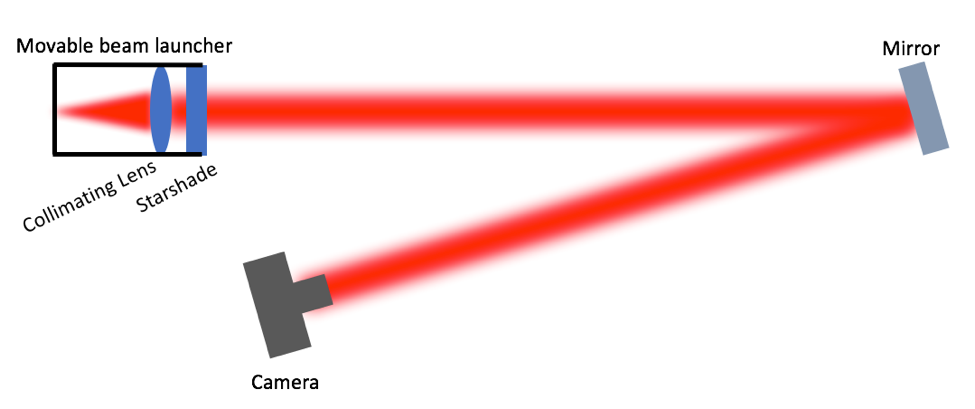}
    \caption{(Left) Schematic of SLATE; movable fiber beam launcher, and fixed fold mirror and camera. }
      \vspace{1.5 cm}
  \end{minipage}
  \hfill
  \begin{minipage}[b]{0.25\textwidth}
    \includegraphics[width=\textwidth]{{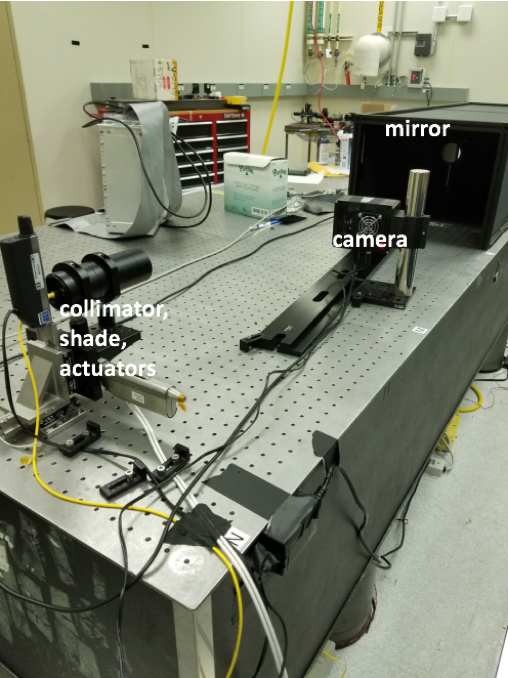}}
    \caption{Image of testbed, partially uncovered.  The camera may be translated along the rail to access different Fresnel numbers.}
  \end{minipage}
  \label{SLATE_images}
\end{figure}

SLATE is a beam launcher and a camera.  The beam launcher consists of an optical fiber, a 100mm doublet collimating lens, and the starshade mask. These optics are small enough to fit on a two-axis stage, creating a movable beam to simulate shear offsets corresponding to starshade motion.  A fold mirror increases the propagation length on the modestly sized optical bench.  The camera takes the place of the low-order wavefront sensor.  Figure \ref{SLATE_images} shows a schematic and image of the testbed.

The camera sees ``pupil'' images, but we do not have a telescope simulator in the beam to create either the pupil of WFIRST, the Zernike phase plate, or any internal recollimating and refocusing optics.  This is by choice, as every optical surface adds noise and complication.  To simulate the telescope pupil, we just mask out the pixels on the camera corresponding to the effective pupil obscuration.  These pixels are not expected to be used in the image matching algorithm in flight either.

SLATE can create optical sensing signals similar to those expected in space, but deviates from a ``perfect'' formation flying lab setup.  A summary of the differences between the test setup and flight expectation is presented in Table \ref{SLATEtable}.

\begin{table}
\centering
\begin{tabular}{lll}
Parameter                     & Flight Expectation                         & SLATE                              \\
\hline\hline\hline
Fresnel number                & 5-7                                        & 4.5                                       \\
Light type                    & broadband starlight (50-100nm filtered)    & 632 nm laser                              \\
Wavefront quality             & \mytilde14 nm wavefront error & \textgreater{500} nm wavefront error  \\
\hline
Camera chip                   & e2v CCD201                                 & SBIG KAF402-me                            \\
Camera read noise             & 2 electrons/pixel/frame                    & 40 electrons/pixel/frame                  \\
Camera dark current           & 1.5e-4 electrons/pixel/second              & 2 electrons/pixel/second                  \\
Camera clock-induced charge   & 0.02 electrons                             & \textless{1} electron                 \\
Camera shutter speeds		& 0.001 - 100+ seconds		     & 0.1-100s\\
Camera flat field calibration & \textless{2}\%         & None                                      \\
\hline
Arago spot FWHM               & 10 pixels / 32x32 pixels                   & 10 pixels/ 32x32 pixels                   \\
Arago spot SNR                & 5/pixel in FWHM                            & 5/pixel in FWHM               \\
\hline \hline
\end{tabular}
\caption{Comparison of optical, detector, and morphological parameters of SLATE with the flight expectation}
\label{SLATEtable}
\end{table}

\subsubsection{Optical considerations}
The contrast of the guiding signal is at 10$^{-3}$ to 10$^{-4}$, which is challenging to achieve optically, but nowhere near as challenging as building a testbed to simulate the optical performance of the starshade at science contrast, at 10$^{-10}$ to 10$^{-11}$.  This is not just due to the optical tolerances, but the amount of light present is much higher, meaning low-noise detectors are not required.  Additionally, effects that are important at the 10$^{-8}$ to 10$^{-11}$ contrast levels are irrelevant for formation flying.  These include edge glint from scattered sunlight, exozodiacal light, and the high likelihood of source confusion from faint background galaxies.\cite{hu2017simulation}  However, attaining shear sensing contrast still requires some care, and we traded off optical tolerances with fidelity to the flight system.  Here, we will describe some differences between the flight system and our testbed.

For the light source, we elected to use a single-mode fiber laser to simulate the starlight, rather than broadband illumination.  Despite the single wavelength (632 nm), the images that a flight pupil sensor will see would be similar, because the sensing wavebands are not particularly broad, at \mytilde10\% optical bandwidth.  Additionally, one side of the band will be much brighter than the other due to the steep wavelength dependence of starshade transmission.

The incident beam on the starshade in flight will be a flat wave of starlight with effectively no aberration.  A flat beam is not an option in this testbed as diffraction from the edges of the optics would overwhelm the faint guiding spot.  We used a beam from a fiber, collimated 100 mm from the starshade, as optical modeling showed the Gaussian shape would marginally affect the spot contrast while eliminating edge diffraction.  Our original choice of a precision asphere to collimate the Gaussian beam failed badly due to significant mid-spatial-frequency errors in the lens.  On the other hand, our optical model indicated that the spherical aberration from an off-the-shelf doublet would marginally affect the guiding signal.

Starshades are meant to be free-floating, which cannot be reproduced in the laboratory.  Mounting them with ``struts'' will cause unacceptable diffraction unless the struts themselves are apodized similarly to the petal edges; this is the approach taken by the experiments at science contrast levels. \cite{kim2015design} In our case, the starshade was manufactured\footnote{by Opto-Line International, Inc.,} by depositing chrome on an optical reference flat.

Beyond static errors like spherical aberration, the biggest optical challenge was mid- to high-spatial frequency error.  We simulated the expected power spectra of the optical surface roughness of the lenses and starshade, and found that operating at Fresnel numbers of F \mytilde 7 would require an RMS surface error of \mytilde 5 nm, which is challenging to achieve without active optical control (for example, interferometer reference flats are typically specified up to 1/20th of a wave).  However, the same simulations indicated that operating closer to F \mytilde4-5 would be achievable with bulk, static optics.  While this is at the lower range of flight Fresnel numbers, it allowed us to validate the simulations without the added complexities of an active wavefront control system.

\subsubsection{Camera parameters}
The flight EMCCD detector for the WFIRST coronagraph, the CCD201 from e2v, is baselined for both the low-order wavefront sensor and science camera.  Unsurprisingly, its performance exceeds the lab detector's by factors ranging from 20 (read noise) to 10,000 (dark current).  As such, attempts to match exposure times and flux levels in the testbed to flight levels would result in a much lower signal-to-noise ratio than that delivered by the EMCCD.  Instead, we adjusted the exposure times and laser power to match the empirically measured signal-to-noise ratio of the spot, which ranges from 3-8 depending on the wavelength.

Another unknown at this time is the pixel resolution of the LOWFS camera, which is expected to be between 16 and 64 pixels across the pupil diameter.  The pixel resolution does not meaningfully affect formation flying performance, provided the sampling is fine enough to properly resolve the spot structure, but not so fine that readout noise overwhelms the signal.  The lab camera's native resolution was about 100 pixels across the pupil, which we digitally interpolated down to 32, for a final output image format of 32x32.

\subsection{Experimental Design}
The experimental setup used signal-to-noise ratios and spot sizes (as fractions of the pupil diameter) determined from the expected flight-like values.  These are listed in Table \ref{SLATEtable}.  The relative scaling between the testbed motion and flight motion (in units of millimeters per cm) were calculated analytically and verified experimentally.  

We experimented with different processing of the camera images, but opted to go with a straightforward ``image minus dark frame'' calibration.  This is due to features of the test camera, which included a fixed bias offset, and confounding factors such as background illumination from neighboring laboratories.  The dark-subtracted frame was then input into the image matching algorithm presented in the Appendix.  While we would have been able to get better performance using more advanced postprocessing, like filtering out optical noise, this minimal level of calibration stays close to the flight algorithm.  It is also expected that in flight, there will be additional error sources like unstable flat fields and charge traps due to cosmic ray damage, which will not be in the optical model.

\begin{figure}[!htb]
  \centering
  \begin{minipage}[b]{0.65\textwidth}
    \includegraphics[width=\textwidth]{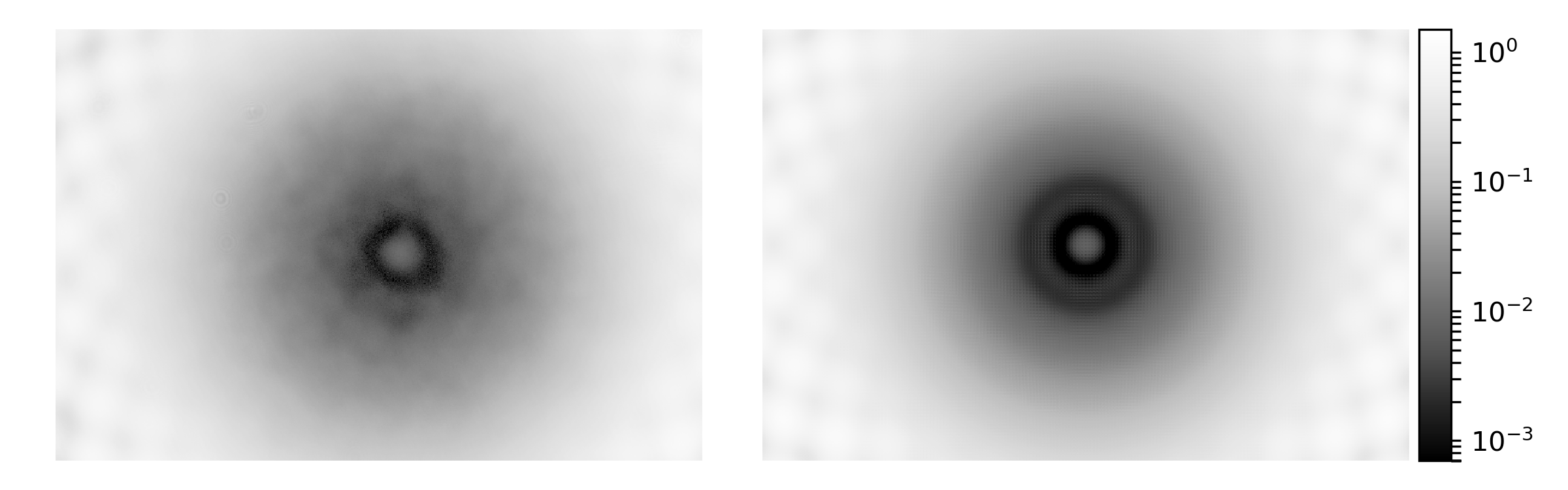}
    \caption{ \label{lab_vs_sim} A comparison of the testbed measurement (left) with the lab simulation (right) on the same logarithmic image intensity scale.  The spot of Arago and other diffraction artifacts are clearly visible.  Note the optical noise in the lab image.}
       \end{minipage}
\end{figure}

\begin{figure}[!htb]
  \centering
  \begin{minipage}[b]{0.65\textwidth}
    \includegraphics[width=\textwidth]{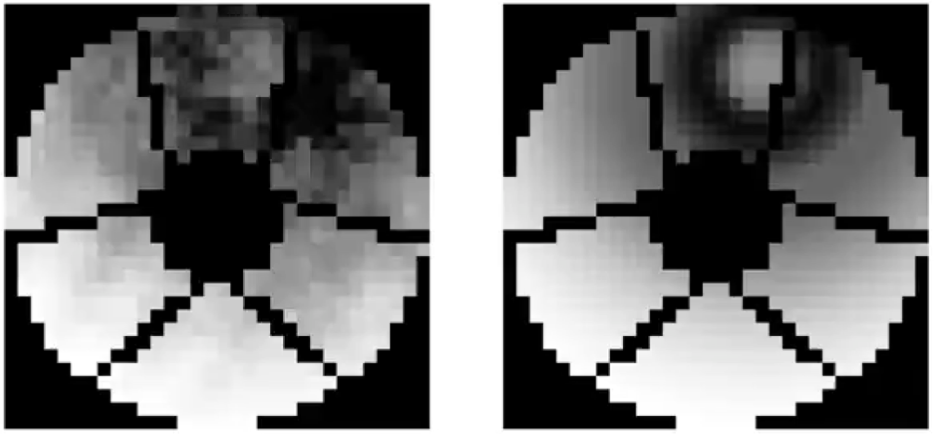}
    \caption{ \label{image_matching} Image matching from the noisy camera image to the model prediction.}
       \end{minipage}
\end{figure}

In order to build the sensor model, we first needed to create an image library for the lab.  We computed the  diffraction pattern on the sensor with models of the lab optics, involving the fiber output beam, collimating lens, and miniature starshade.  Obviously, sizes and distances are different as well, at \mytilde3 meter separations and a \mytilde6mm starshade rather than \mytilde40,000 km separations and a 26 meter starshade.  An example of the computed and measured diffraction pattern, before adding the pupil and binning, is shown in Figure \ref{lab_vs_sim} (Figure \ref{image_matching} shows the 32x32 version with the pupil overlaid).  While our contrast measurements were consistent to 25\% of expectation, absolute values of contrast are not relevant because the images are normalized before being matched.

We determined the correct laser drive voltage by empirically measuring the signal-to-noise ratio of the pixels (after being binned to the LOWFs plate scale) as approximated by SNR = mean/(standard deviation).  We matched the empirical signal-to-noise ratio to that expected in space from stars fainter than 8th magnitude at a typical spot size.  The exact nature of the noise will change between the flight detector and SLATE; the former will be almost purely Poissonian, while the latter includes Poisson, readout, and dark current noise.  It would be in principle possible to independently characterize the different SLATE detector noise sources, and their combined distributions, but we opted to use the empirical SNR instead.  While we did not analyze the exact form of the noise statistics, it is possible that some variation in our results is due to these subtle effects.

While the actuator encoder values could be used for open loop positioning, they had a slight tilt with respect to the optical axis and some backlash.  Rather than try to calibrate these imperfections, we determined the position of the starshade directly from the camera images and ran an acquisition loop to go to the preset  grid points, spaced apart by 30 cm (effective).  To minimize errors during acquisition due to optical and detector noise, the laser was turned to a bright level such that the image matching always returned the same result.  Then the laser was turned down to the ``science intensity'' and hundreds of frames were taken at those flux levels.  At each grid point, the images were matched to the library, and the corresponding positions were used to generate error ellipses shown in the next section.  An example of a single camera image and its matched model are shown in Figure \ref{image_matching}.

\subsection{Results}
\begin{figure}[!htb]
  \centering
  \begin{minipage}[b]{0.75\textwidth}
    \includegraphics[width=\textwidth]{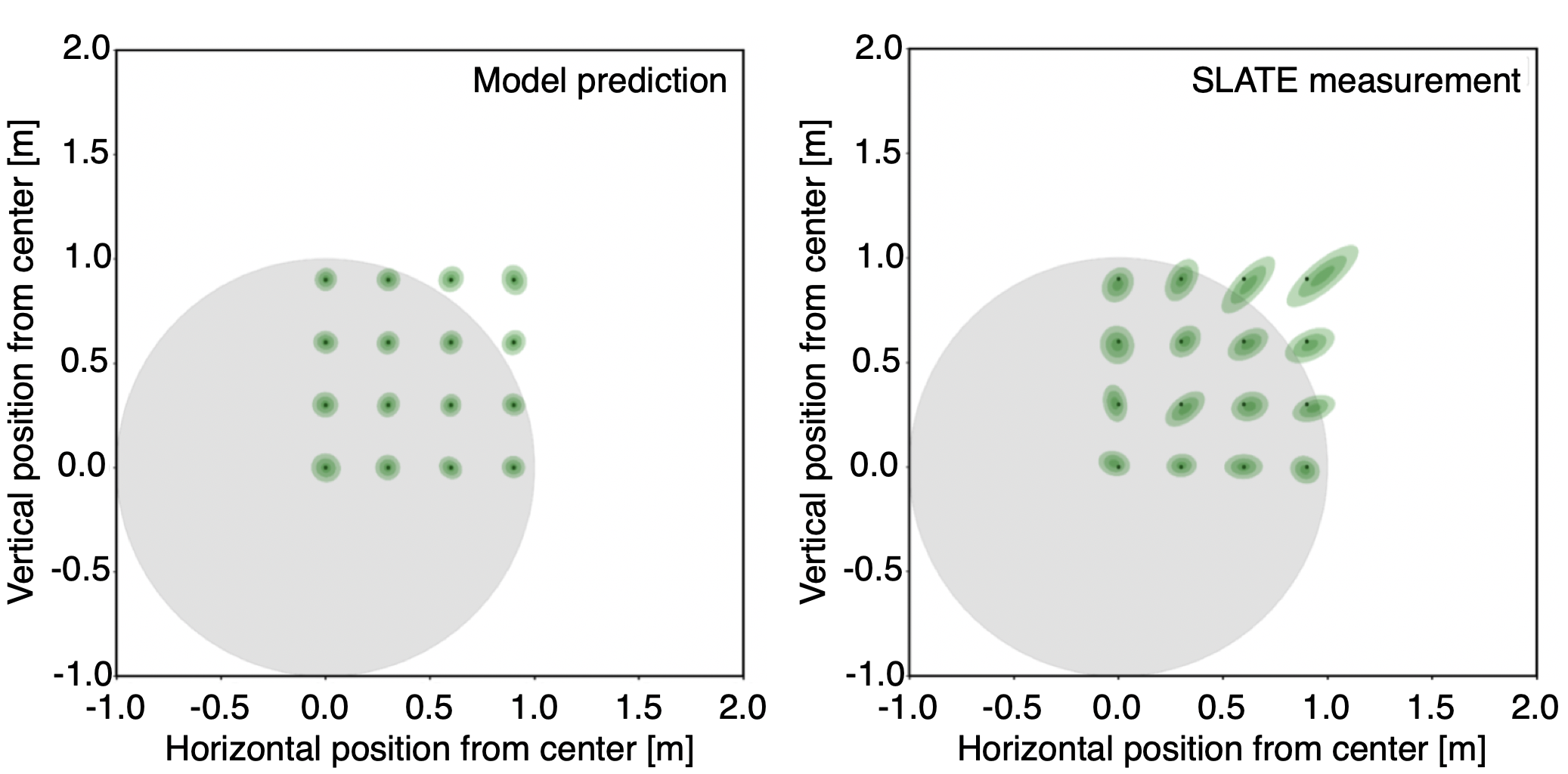}
    \caption{ \label{slate_vs_sim} Results of sensing precision of SLATE (right) and a model of SLATE at the same flux levels (left).}
       \end{minipage}
\end{figure}

From the empirical data covariance matrix at each position, we generate error ellipses showing, 1, 2, and 3$\sigma$ contours.  These results were consistent with numerical expectations within a factor of \mytilde50\%, as shown in Table \ref{sim_vs_slate_table}.    Plots of the results are presented in Figure \ref{slate_vs_sim}.  The delivered sensing precision, which was obtained at much lower signal level than expected in flight, is still well within the tolerance specified for formation flight, and a companion work will demonstrate robust control even with errors far larger than what the sensor can deliver.

\begin{table}[htp]
\caption{\label{sim_vs_slate_table}Comparison between numerical simulations and actual testbed performance of the worst and median 3-$\sigma$ sensor precision}
\begin{center}
\begin{tabular}{| l |c | r |}
Simulation & 3$\sigma$ precision, worst & 6.7 cm\\
SLATE & 3$\sigma$ precision, worst & 10.2 cm \\

Simulation& 3$\sigma$ precision, median & 4.0 cm \\
SLATE & 3$\sigma$ precision, median & 6.2 cm \\
\end{tabular}
\end{center}
\label{default}
\end{table}%

The primary reason for the worse performance in the lab than from numerical simulation is optical noise; that is, blobs of bright light forming structures in the shadow, that are not expected in the CGI flight optical system. This noise is created by light scattered by the imperfect optics.  The extra optical noise leads to both statistical and systematic errors.  The statistical errors are due to the combination of Poisson, dark, and readout noise; the systematic errors are due to the matching algorithm biasing towards the scattered light structures.  (These systematic errors are visible in Figure \ref{slate_vs_sim} as slight shifts in the midpoint of the error ellipses compared to the setpoints).  

Errors in the camera also contribute.  The dark level in the camera drifts continuously, and while we always took a background frame before a science frame, the noise on top of these frames can appear as a changing noise gradient from one side of the detector to the other.  Another issue was flat-field correction: we did not solve for a flat-field on the camera and thus differences in per-pixel gain can create a spatially dependent systematic error signal.  These errors will also be present in flight to some degree, as cosmic ray damage begins to affect the pixels in the detector.

\section{Conclusions}
We have presented a lateral sensing scheme appropriate for the challenging task of starshade formation flying, where two spacecraft must be aligned to a precision of 1 meter at distances of 20,000-80,000 km.  The sensing scheme measures the position of the classical ``Arago spot'' from light diffracting around the edges of the starshade using an internal pupil sensor on the telescope.  This light, which is outside the wavelengths of scientific interest, bright enough to provide a robust sensing signal.  The precision of this sensing scheme is just a few centimeters in shear for star brightnesses of 8-10 $V$ magnitudes, which is fainter than any of the expected target stars by factors of $>$10.  The performance of this sensor shows good agreement when compared to analytical calculations, detailed numerical simulations, and laboratory experiments.

No additional hardware is needed to implement this sensor beyond a pupil imager in the telescope.  This is already present in the case of the WFIRST coronagraph instrument, where it is used as an \textit{internal} wavefront sensor in coordination with a Zernike spot.  In the case of future missions like mDot,\cite{d2019system} HabEx,\cite{mennesson2016habitable} or LUVOIR \cite{luvoir2018luvoir}, pupil sensors are expected to be present as well.  As such, they can support accommodation for future starshade rendezvous missions.

A companion paper (Flinois et al., in preparation)\cite{flinois_inprep} will introduce a control scheme that can easily provide enough fidelity to keep the starshade and telescope aligned to the 1 meter necessary for imaging extrasolar planets.  Such a control scheme has a high level of efficiency and an ability to execute nearly optimal trajectories in the differential gravity of L2, with minimal interruptions in science operations for trajectory correction maneuvers.  As such, we have high confidence that the formation flying problem, which was initially considered a major challenge in implementing a starshade, can be solved.

\subsection*{Disclosures}
The authors declare no financial or other conflicts of interest in this publication.

\acknowledgments 
The authors wish to thank the referees for constructive reports that improved the quality of this paper.  MB thanks Kendra Short, Phil Willems, Doug Lisman, and Charley Noecker for insights and guidance.  The authors thank Opto-Line International, Inc. for their workmanship and timely delivery of our mini starshades. This work was performed at the Jet Propulsion Laboratory, California Institute of Technology, under contract with the National Aeronautics and Space Administration. (C) 2019. All rights reserved.

\appendix
\section{Storage and computational requirements for formation sensing algorithm}
The baselined flight computer for the coronagraph instrument on WFIRST is the LEON4 processor\cite{andersson2017leon} from Cobham Gaisler.  Internal testing of the processor report a maximum performance of 800 MFLOPS, with an ``effective'' performance (including all low-level overheads) estimated at 76 MFLOPS.  The available memory allocated to the coronagraph instrument is about 66 GB.

\subsection{Library size and storage requirements}
For formation flying, the control region is 1 meter in radius.  It is expected the image library size will be larger, to accommodate some margin for error and initial acquisition.  We assume a 3x3 meter library, spaced at 2 cm, for (3m/2cm)$^2$ = 22,500 images.  For 32x32 images stored as 16 bits/pixel, we find a total space requirement of about 22,500*32*32 pixels*16 bits/pixel = 368 Mb = 46 MB.  Assuming one library is used for each of the three science bands, this leads to a total space requirement of about 150 MB for formation flying, or about 0.2\% of the total storage space allocation of 66 GB.

\subsection{Computational requirements}
Here we describe a brute-force implementation of the image matching algorithm.  Let \textit{n} be the number of pixels used in the pupil sensor; for example $n=1024$ for the $32$x$32$ format considered earlier.  Let $m$ be the number of images in the library.  Table \ref{FLOPS} outlines the steps in the shear-sensing algorithm with their associated floating point cost.

\begin{table}[htp]
\caption{Floating point cost for shear-sensing algorithm}
\begin{center}
\begin{tabular}{| l |c| r|}
Operation & Algebraic description &Floating-point operations\\
\hline 
Sum raw image intensities 				& 	$\sum I $					&		$n-1$\\
Calculate the mean intensity 				& $\left<I\right>= \sum I/n$		&		1\\
Divide raw image by mean intensity 			& $I_\mu= I/\left<I\right>$			& 		$n$\\
Subtract result from each image library image 	& $d_{x,y} = L_{x,y}-I_\mu $		& 		$nm$\\
Square result 							& $ d_{x,y}  \rightarrow d_{x,y}^2$	&		$nm$\\
Sum results							& $e^2_{x,y} = \sum d_{x,y}^2$		&		$(n-1)m$\\
Find minimum error using binary search		& $ \mathrm{min} [e_{x,y}^2]$		&		$m-1$ \\
\hline
Total cost								&		  					&   		$3nm +2n + 1$\\

\end{tabular}
\end{center}
\label{FLOPS}
\end{table}%

To convert to FLOPS, we assume each low-level mathematical operation (addition, multiplication, etc) costs 1 FLOP.  For the 22,500 images per library, a full search each second would cost about 70 MFLOPS, which is just within the capabilities of the LEON4.  However, after the initial position fix (using perhaps a much coarser library spaced at 10 cm), a search over the entire image library will not be necessary, since the starshade does not move  fast.  With a maximum expected speed of 2 cm/s, in one second of motion only about 25 images would need to be searched.  Being more conservative, one could search a 10 cm radius around the previous fix with a sub-library with less than 100 images, which would consume less than 0.5 MFLOPS, or about 1\% of the 76 MFLOPS capabilities of the LEON4.  The entire calculation would take about 6 ms, contributing 0.6\% to the 1 second sensing cadence.

\bibliography{report}   
\bibliographystyle{spiejour}   

\vspace{1ex}
\noindent Biographies and photographs of the other authors are not available.

\listoffigures
\listoftables

\end{spacing}
\end{document}